\documentclass[journal]{article}

\usepackage{arxiv}
\usepackage{xcolor,soul,framed} 

\colorlet{shadecolor}{yellow}
\usepackage[pdftex]{graphicx}
\graphicspath{{../pdf/}{../jpeg/}}
\DeclareGraphicsExtensions{.pdf,.jpeg,.pdf}

\usepackage[cmex10]{amsmath}
\usepackage{array}
\usepackage{mdwmath}
\usepackage{mdwtab}
\usepackage{eqparbox}
\usepackage{url}
\usepackage{subcaption,enumitem,natbib}
\usepackage{booktabs} 

\title{Bridge Structural Health Monitoring using Asynchronous Mobile Sensing Data}

\author{
  Soheil Sadeghi Eshkevari \\
  Lehigh University\\
  \texttt{ses516@lehigh.edu} \\
  \And
  Liam Cronin \\
  Lehigh University \\
  \texttt{lmc219@lehigh.edu} \\
  \And
  Shamim N. Pakzad \\
  Lehigh University\\
  \texttt{pakzad@lehigh.edu} \\
  \And
  Thomas J. Matarazzo \\
  MIT Senseable City Lab \\
  \texttt{tomjmat@mit.edu} \\
  }

\begin{document}
\maketitle

\begin{abstract}
This study presents a flexible approach for bridge modal identification using smartphone data collected by a large pool of passing vehicles. With each trip of a mobile sensor, the spatio-temporal response of the bridge is sampled, plus various sources of noise, e.g., vehicle dynamics, environmental effects, and road profile. This paper provides further evidence to support the hypothesis that through trip aggregation, such noise effects can be mitigated and the true bridge dynamics are exhibited. In this study, the continuous wavelet transform is applied to each trip, and the results are combined to estimate the structural modal response of the bridge. The Crowdsourced Modal Identification using Continuous Wavelets (CMICW) method is presented and validated in an experimental setting. In summary, the method successfully identifies natural frequencies and absolute mode shapes of a bridge with high accuracy. Notably, these results are the first to extract torsional mode shape information from mobile sensor data. Moreover, the influence of vehicle speed on the estimation accuracy is investigated. Finally, a hybrid simulation framework is proposed to account for the vehicle dynamics within the raw mobile sensing data. The proposed method is successful in removing vehicle dynamic effects and identifying modal properties. These results contribute to the growing body of knowledge on the practice of mobile crowdsensing for physical properties of transportation infrastructure.

\end{abstract}

\section{Introduction}\label{sec:introduction}

Infrastructure is a vital component of transportation systems and a vibrant economy. In many countries, monitoring and maintenance of existing infrastructure have turned into a significant concern for urban planners and decision makers, especially in the US, Europe, and Asia \citep{lancefield2017thousands, willsher2018bridges, Perez2018after}. Nearly one of every nine bridges in the US is structurally deficient according to ASCE \citep{american20172017}. The global state of infrastructure indicates a demand for large-scale health monitoring solutions with reduced setup costs that can rapidly produce information on existing mechanical properties, damage location, and extent of deterioration in bridges. Among available health monitoring methods, vibration-based systems have become popular as data acquisition is relatively inexpensive and the resulting data is an asset to a bridge management system.

\subsection{Crowdsourcing for Urban Transportation Sensing}

Through widespread presence of smart devices, datastreams are available that are primarily collected for personalized, comfort-intended tasks, such as adaptive screen lighting or navigation, but they can be further analyzed for widely-impacted socio-economical secondary objectives. Smartphones are equipped with numerous sensors for motion measurement and have an ability to continuously record data at high rates. Of course, in comparison to more dedicated instruments, smartphone sensors are generally less precise and are expected to produce noisier data; nevertheless, the datasets produced by phones can be extremely large and include simultaneous measurements from dozens of sensing channels. In some cases, the vast size of these datasets can help overcome shortcomings in data quality. Recent research explored the possibility of using such data from vehicle networks for urban planning and transportation applications. A real-time framework was introduced that is able to create urban dynamic maps using smartphone data, such as traffic conditions and pedestrian movements in Rome \citep{calabrese2010real}. \cite{wang2012understanding} by combining GIS and smartphone data detected the patterns of road usage and origins of the cars, which are applicable for transportation planning. A crowdsourced-based algorithm was proposed by \cite{shin2015urban} for automated transportation mode detection using smartphone sensing data including acceleration. Other applications such as road incident reporting and real-time parking space information based on crowdsourcing data from drivers are proposed as well  \citep{wang2017crowdsensing,shi2018parkcrowd}. A review study on smartphone data crowdsourcing for vehicle positioning and drivers' behavior monitoring is conducted by \cite{kanarachos2018smartphones}. The study elaborated that smartphones are advantageous for crowdsourcing application due to the market penetration, \textit{Internet of Things} connectivity, and data sharing capabilities.\par

Smartphones as sensors is an emerging paradigm which is not limited to traffic monitoring and has recently been applied towards health monitoring of transportation infrastructure. The state of practice in vibration based structural health monitoring (SHM) is to utilize dedicated accelerometers at fixed locations and periodically monitor the output. However, mobile sensors have offered a great potential to dramatically overhaul this conventional paradigm. Some studies proposed smartphone acceleration and GPS data for real-time localization and characterization of road bumps from vehicle networks \citep{mednis2011real,kumar2016community,mukherjee2016characterisation}. More recently, a vision-based approach for road condition assessment based on images from dashboard-mounted smartphones has been proposed and experimentally tested \citep{maeda2018road}. These successful applications encourage SHM community to develop crowdsensing-based methodologies for bridges.

\subsection{Bridge Health Monitoring}

Over the past twenty years, there has been extensive research on wireless sensor networks for structural health monitoring (SHM) and nondestructive testing applications. These sensor networks are economical, easy to implement and can support denser networks when compared with traditional wired systems \citep{pakzad2008design,lynch2006summary,kim2007health,harms2010structural,hackmann2013cyber}. There is a plethora of important studies that successfully implemented dense wireless sensor networks on real-world bridges and have shaped the SHM field \citep{kim2007health,lynch2005validation}. With the emergence of wireless sensors, \textit{spatial coverage} and \textit{spatial density} have become increasingly important features of any sensor network as they broadly affect the veracity of key applications, e.g. damage localization, model updating, etc. The primary approach for enhancing spatial density of the identified modal properties is simply to design sensor networks with additional sensors \citep{lynch2006summary,kim2007health}; the drawbacks of this approach are an increased complexity in data acquisition and transmission systems and higher setup costs.

\subsubsection{Mobile Sensing for Bridge Health Monitoring}
Over the years, the capabilities of mobile sensor networks in SHM have been demonstrated both in theory and in practice. Mobile sensors have outstanding benefits: (1) the data can contain rich spatial information; (2) ubiquitous smartphones are great candidates for mobile sensors. (3) when combined with vehicle networks or fleets, mobile sensors routinely scane transportation infrastructure.

There has been progress in developing general purpose modal identification methods for mobile sensor data that are not necessarily tied to vehicles. The structural identification using expectation maximization (STRIDEX) \citep{matarazzo2018scalable} utilized data from multiple mobile sensors for comprehensive modal identification (frequencies, damping ratios, and determine mode shapes) and further quantified the superior spatial information that can be produced with mobile sensor networks. With only two mobile sensors, STRIDEX identified the fundamental mode shape with $248$ points - that is, two mobile sensors captured spatial information that was comparable to 120  fixed sensors \citep{matarazzo2018scalable}. The modal identification through matrix completion (MIMC) method \citep{eshkevari2020signal,eshkevari2020mimc} is a sparse sensing approach for processing large-scale sensing scenarios (mobile, fixed, or hybrid). A benefit of this method is that it can handle data from a very large number of moving sensors with arbitrary bridge trajectories. \par

Simultaneously, ``indirect monitoring'' research has considered measurements of vehicle vibrations collected while passing over bridges, which are subject to vehicle-bridge interaction  \citep{yang2004extracting,lin2005use,yang2009extracting,siringoringo2012estimating,gonzalez2012identification}. These studies have mostly considered one vehicle, simplistic traffic cases, and have been limited to partial modal identification of the bridge, e.g., frequencies only, damping ratios only. In these studies, a coupled dynamical system including both the vehicle and the bridge is usually developed, which is subject to a controlled load from the moving vehicle. A recent study by \cite{sadeghi2020simplified} showed that in certain traffic scenarios, the vehicle dynamic response is nearly decoupled from the bridge system when the collective traffic loads on the bridge are effectively random.  \par 

\subsubsection{Crowdsensing for Bridge Health Monitoring}

The previous studies focused on mobile sensing for bridge health monitoring, yet did not strictly analyze smartphones for crowdsensing. Early studies with on-board smartphone acceleration sensors for structural health monitoring applications provided preliminary validation of smartphone sensor accuracy in a laboratory setting \citep{yu2015initial,feng2015citizen,ozer2015citizen}. \par
An advantage of smartphones is that they contain other useful sensors such as GPS, gyroscopes, magnetometers, etc. which facilitates sensor data fusion and cross-validation among reference sensors \citep{guzman2019gps}. For example the integration of Intelligent Transportation Systems (ITS) including traffic video camera and bridge structure instrumentation with dedicated sensor data was suggested in order to enhance the estimation accuracy for structural health monitoring applications \citep{gandhi2007video,khan2016integration}. A particular application of interest is of course crowdsourcing smartphone data to assess the condition of existing infrastructure, e.g., bridges. \cite{matarazzo2017smartphone} analyzed call detail record (CDR) data to estimate the number of smartphones that cross the Harvard bridge each month. Such estimates help quantify the crowdsensed data potential for specific urban infrastructure based on human mobility patterns. A later study aggregated smartphone data collected from forty-two vehicle trips over the Harvard bridge and extracted consistent indicators of the first three modal frequencies \citep{matarazzo2018crowdsensing}. This study was the first to support the hypothesis that smartphone data, collected within vehicles
passing over a bridge, can be used to detect several bridge modal frequencies. Other studies involving mobile smartphone sensors have developed methods that can successfully track damage-sensitive features in a laboratory setting \citep{mei2018crowdsourcing,liu2020diagnosis}.  \par


A recent study at MIT's Senseable City Lab proposed a wavelet-based statistical methodology which aggregates smartphone vehicle-trip data to estimate the most probable modal frequencies (MPMFs) of a bridge. The study considered a large number of controlled and uncontrolled vehicle trips and successfully estimated the first three modal frequencies of the Golden Gate Bridge using the proposed framework. The method uses synchrosqueezed wavelet transformations to produce a stack of spatial-frequency maps whose local maxima represent bridge modal properties. The primary advantage of the proposed method is that it does not require synchronized sensing. In other words, sensor data aggregation is performed regardless of the sensing time for individual scans. This paper builds on the same strategy by proposing a new wavelet-based methodology that is able to identify the modal frequencies and absolute mode shapes from crowdsourced mobile smartphone data.

\subsection{Spatio-temporal transformations}


In this paper, the primary objective is to estimate natural frequencies and absolute values of mode shapes by crowdsourcing mobile smartphone data. As explained in \citep{eshkevari2020mimc} and \citep{matarazzo2018crowdsensing}, bridge dynamics produce a spatio-temporal response which consists of modal responses. For instance, a bridge under a uniform ambient load produces a stationary signal at the fixed locations (since the signal is not a function of space), while a mobile sensor that drives by a bridge collects a nonstationary signal. Time-frequency transformations are designed to represent a nonstationary or nonlinear signal by a map of its time-dependent frequencies. Short-time Fourier transform (STFT) is a common approach that stacks FFTs (fast Fourier transforms) of sliding windows of a signal to create a 2D time-frequency representation (TFR) \citep{zhang2012damage,malekjafarian2014identification}. Continuous wavelet transform (CWT) is numerically more stable and computationally less expensive. In CWT, a mother wavelet is applied on the signal to detect associated frequency contents and then repeatedly scaled up to represent lower frequency contents. The method has been extensively used for damage detection and modal identification. A modified version of wavelet-based methods is synchrosqueezed wavelet transform (WSST), which was used in recent study at MIT's Senseable City Lab, yielding promising results. This algorithm is designed to remove the frequency smearing effect of CWT by squeezing adjacent frequency ridges into the most powerful band.

\begin{figure}[!h]
    \centering
    \includegraphics[width=80mm]{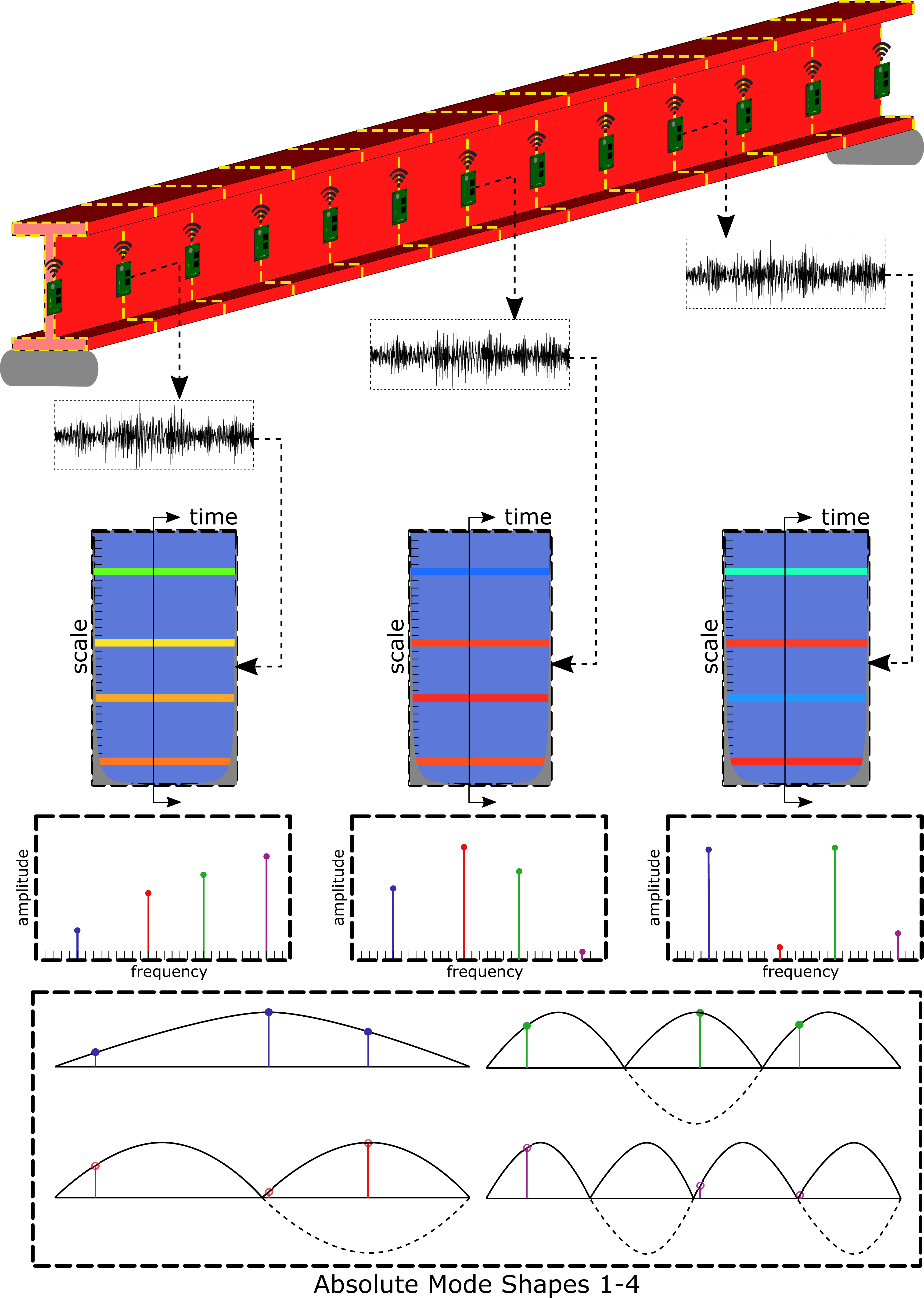}
    \caption{Spatial variation of CWTs: the simply-supported beam is equipped with fixed sensors. The CWT plots of using signals from different fixed sensors are shown. In each CWT plot, the modal bands have amplitudes that are proportional to the spatial amplitude of the natural modes. Frequency bands are consistently color-coded with respect to their intensity. By stacking these CWTs in spatial order, absolute mode shapes are identified.}
    \label{fig:CWT_proof}
\end{figure}

The continuous wavelet transform of a signal collected by a mobile sensor creates a 2D time-frequency representation, in which the time axis is interchangeable with location since $x=f(v,t)$ (where, $x$ is the location, $v$ is the vehicle speed, and $t$ is the time; when vehicle speed is constant, $x=v\times t$). If the structural properties remain linear, this 2D representation includes ridges at the natural frequencies of the bridge. The magnitudes of the ridges are directly dependent on the location. For instance, for a simple beam, the ridge associated with the second mode has zero magnitude at the midspan since the second mode shape has zero magnitude there. Expanding the same argument, it is deduced that the magnitudes of each ridge are directly associated with the absolute amplitudes of the corresponding mode shape at that location. The idea is better explained in Figure \ref{fig:CWT_proof}. In this figure, stationary sensors collect signals and their CWTs are calculated (the complex Morlet wavelet transform used in this example is equivalent to the Fourier transform \citep{bentley1994wavelet}). Among all CWTs, the outstanding frequencies are identical, but their magnitudes vary between locations. By stacking these frequency representations in their spatial order, the envelopes of each modal frequency represents the absolute natural mode shapes. 

\subsection{Motivations and contributions}

Previous studies on mobile sensing have proposed effective solutions for a variety of mobile sensing scenarios \citep{matarazzo2018scalable,eshkevari2020mimc}. Despite their appealing performance and high accuracy, they are limited to scenarios in which higher quality accelerometers (not smartphones) are available. This paper, proposes crowdsourced modal identification using continuous wavelets (CMICW) which is designed to \textit{process} smartphone data as a ``collection'' rather than, one-by-one. The method is able to estimate both natural frequencies and the absolute natural mode shapes (vertical and torsional). Notably, CMICW is the first method with an ability to extract torsional mode shape information from mobile sensor data, (in a generic sense, not only smartphone data). The method also has beneficial computational features as the estimation process is scalable with respect to the number of data sets. \par
The approach relies on complex Morlet wavelet transform to convert individual mobile scans into a 2D map of bridge location versus frequency. By crowdsourcing a large number of one-way scans and producing a library of 2D CWT plots, aggregation minimizes estimation biases through canceling out trivial ridges and intensifying consistent ridges, i.e., absolute natural mode shapes. In such a crowdsourcing scenario, the number of scans in the aggregation pool is the controlling parameter and directly influences the confidence and accuracy of the estimations. In Section \ref{sec:method}, an overview of the designed pipeline is introduced and its components are justified. Section \ref{SEC:Experimental} will present the procedure and results of experimental case studies consisting of a laboratory-scale beam setup. Smartphones moving at various speeds collect vibration data when the beam is subjected to trains of random impulsive loads. The effect of sensors speed variations is also demonstrated. Moreover, a hybrid simulation procedure is designed to integrate vehicle suspensions effect to sensors measurements and the effect is investigated. In section \ref{sec:dicussion}, results are further discussed and the study is summarized. 


\section{Methodology}\label{sec:method}

\begin{figure}[!h]
    \centering
    \includegraphics[width=80mm]{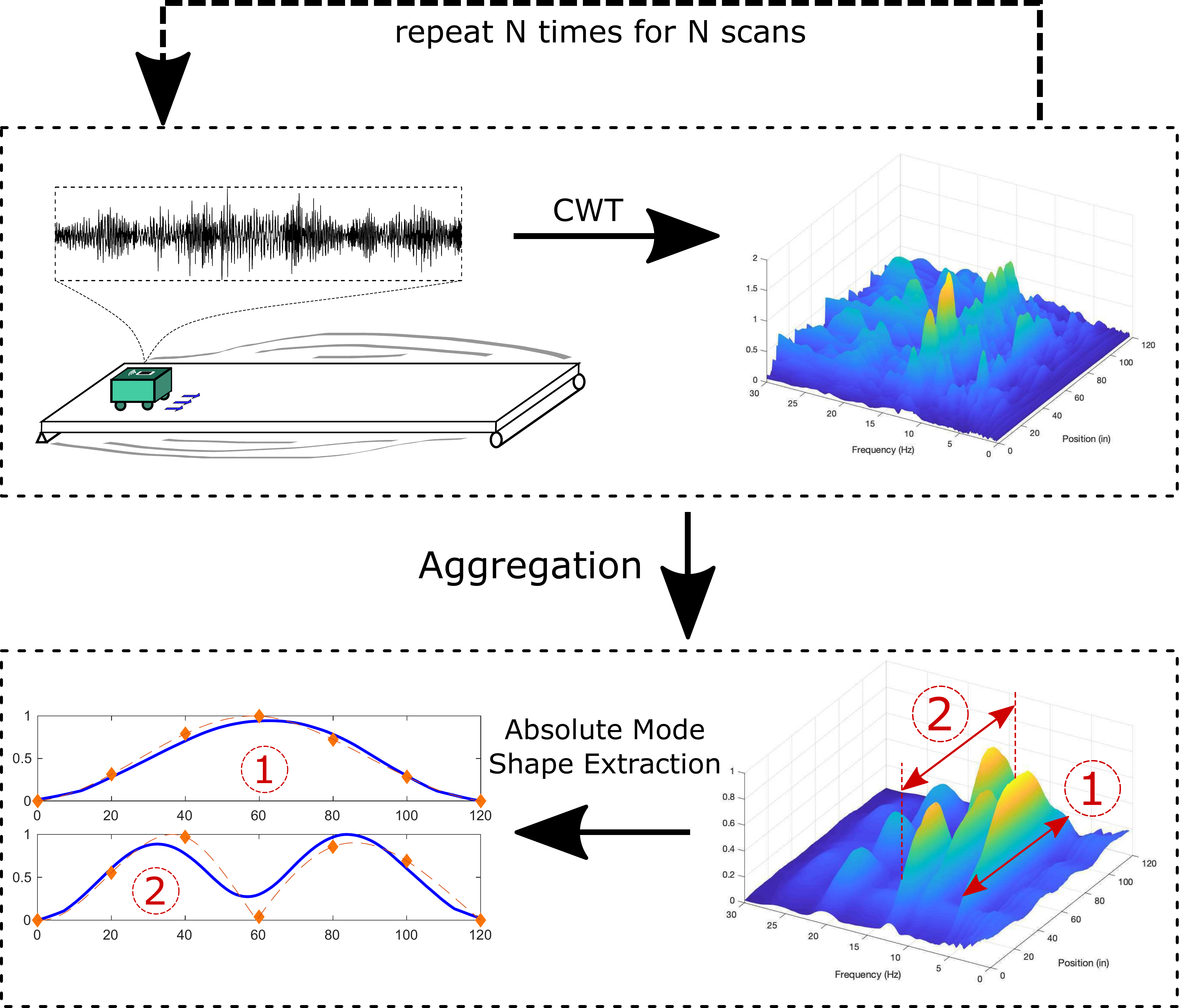}
    \caption{Crowdsourcing schematic for CMICW: each one-way vehicle scan collects a bridge response signal in a mobile fashion. The CWT of each signal includes ridges associated with natural modes that are severely mixed with noises (ambient ridges). However, once a large pool of CWT maps from different signals is collected, the average CWT map yields clear ridges on natural modes with minimal noisy contents. Note that each scan is independent to the rest (temporally and spatially).}
\label{fig:crowdsourcing}
\end{figure}

\begin{figure*}[!h]
    \centering
    \includegraphics[width=150mm]{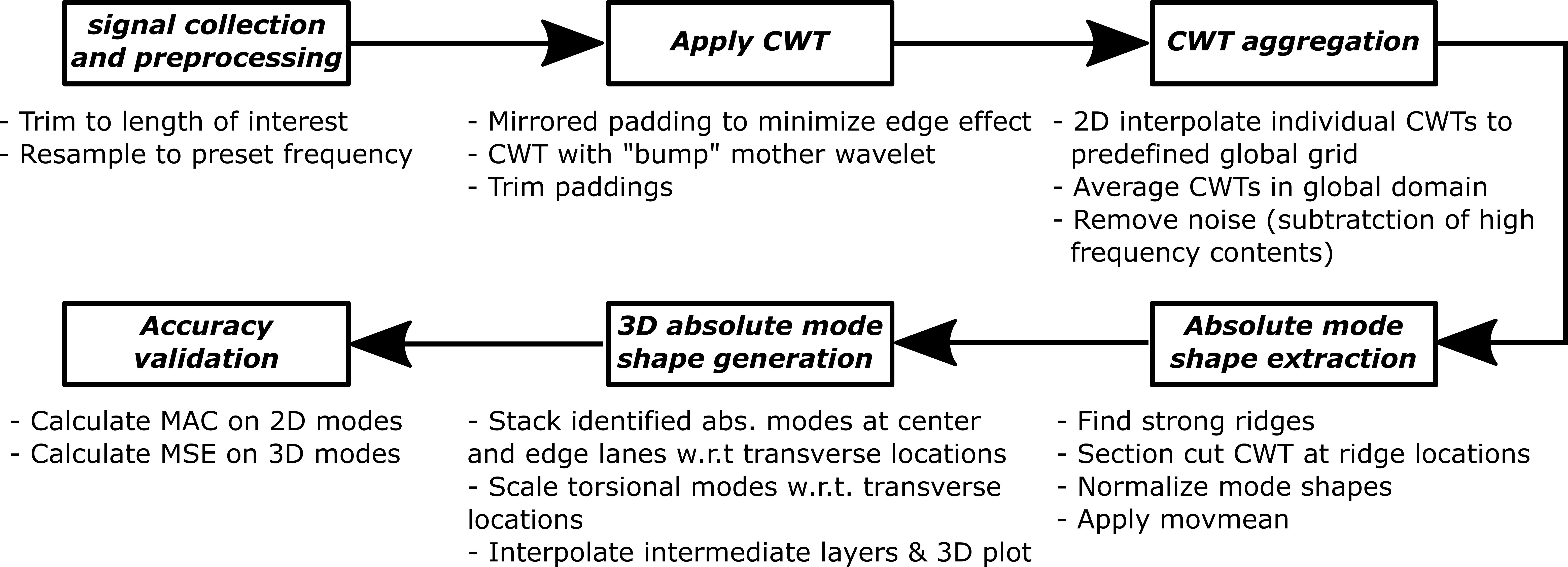}
    \caption{Detailed components of the CMICW pipeline.}
\label{fig:pipeline}
\end{figure*}

A schematic plan of the CMICW pipeline is shown in Figure \ref{fig:crowdsourcing}. In this figure, the starting point is data collection using passerby sensors over the bridge. Each individual scan is fully independent and can happen with or without the presence of other scanning devices. The mobile sensor collects a time history of the bridge vibration responses at different time-location coordinates, depending on the speed of the vehicle. In the next step, this signal is transformed into a time-frequency domain using CWT. For a sensor with constant speed, the time axis can be linearly scaled to a bridge location axis. The CWT yields a 2D map of instantaneous frequencies at different bridge locations. For an individual signal, the 2D frequency map is highly contaminated by trivial contents caused by ambient loads, measurement noises, and stochastic events. Therefore, the process of collecting signals and applying CWT is repeated in order to populate a crowdsourced pool of CWT maps. Once the pool is sufficiently large, an average of 2D maps converges to the frequency-location map of the bridge, i.e., natural frequencies and absolute mode shapes. However, since mobile sensors may have different speeds while sensing, the CWT maps have different resolutions and may not be readily consistent in size (dimensions of the CWT map depends on the length of the signal, which itself relies on the vehicle speed). Hence, the CWTs are 2D interpolated to a predefined global grid. Despite its desired spatio-temporal representation, CWT results in distorted values near the ends of the signal, widely known as the edge effect \citep{montanari2015padding}. To minimize the distortion, the proposed approach from \citep{montanari2015padding} is adopted. After aggregation, the 2D map shows strong ridges on the natural frequencies and minimal noise bed elsewhere, since the noisy contents cancel out each other. These intensified ridges are then automatically detected using a peak-picking algorithm and the aggregated map is section-cut at those frequencies. The final product of the pipeline is the identified absolute mode shapes, which have very high spatial resolution (final output of the procedure shown in Figure \ref{fig:crowdsourcing}). The detailed components of the pipeline are presented in Figure \ref{fig:pipeline}.

In addition to the vertical modes, the method is also capable of reconstructing torsional modes and creating 3D representations. For this purpose, the pipeline is implemented separately for different lanes of the bridge. For the torsional modes, the values of the identified ridges vary between lanes. For instance, torsional modes have zero amplitude when centerline of a beam is considered while the edges reflect these modes with maximum amplitude. Extracting modal ridges in different lanes and placing them in parallel with the same order as the actual lanes construct a 3D representation of the torsional mode shapes. 

\section{Experimental Case Study}\label{SEC:Experimental}

This section presents the setup of the laboratory-scale experiment and the results obtained from implementing the proposed method for modal identification along with validating estimations using fixed sensors as a benchmark. The test mimics a bridge subjected to random ambient load, monitored by moving sensors at random time intervals that are independent. Results show that the procedure successfully identifies three vertical modes and two torsional modes using mobile smarthphone data.\par

\subsection{Test setup}\label{Setup}

\begin{figure}[!h]
    \centering
    \includegraphics[width=85mm]{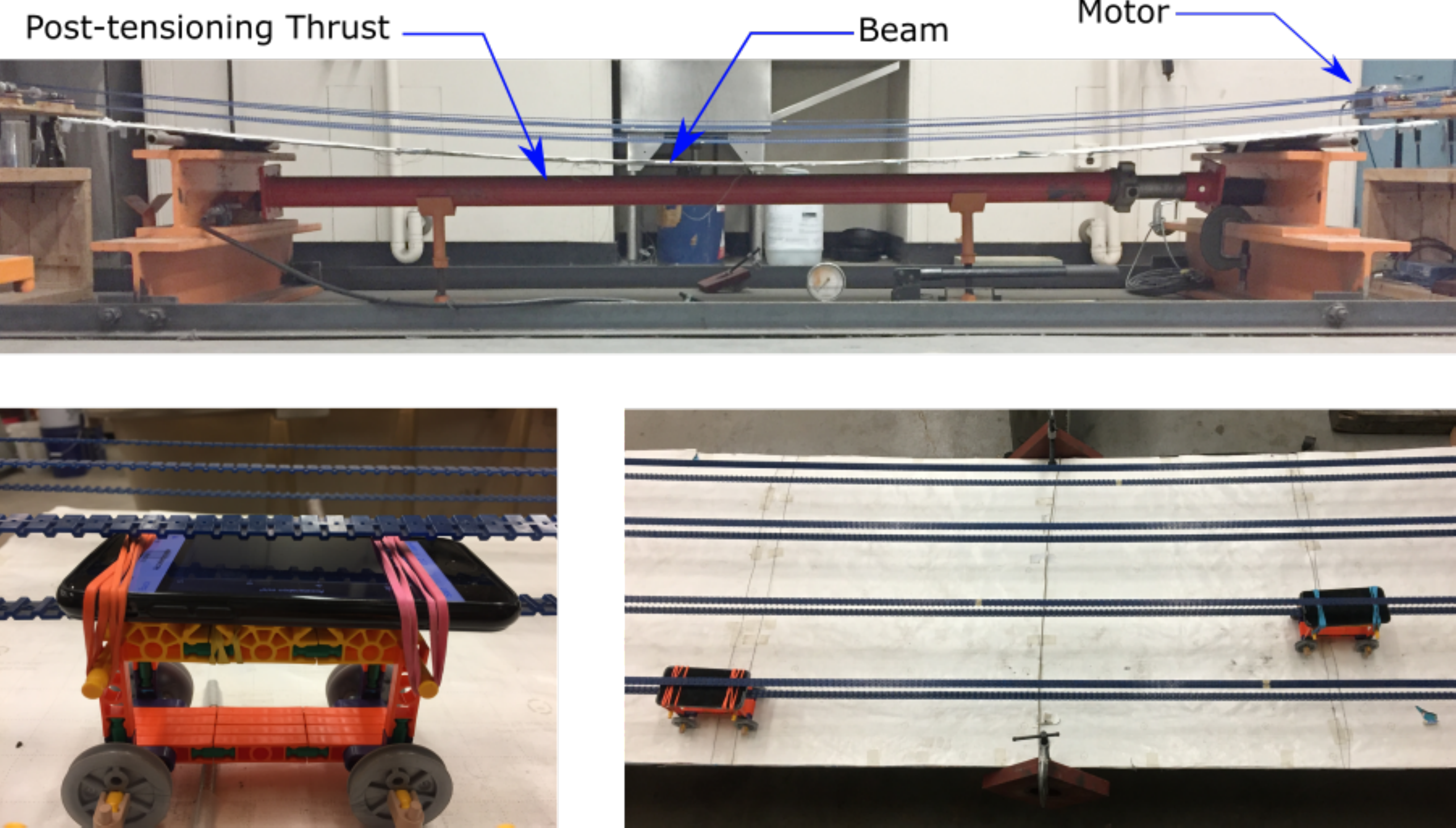}
    \caption{Layout of the test setup: (top) side view of the beam, (bottom-left) individual mobile sensing node, and (bottom-right) mobile sensors in motion.}
    \label{fig:actualsetup}
\end{figure}

The test setup consists of a steel beam that is $3.66$m long, $0.635$m wide and $6.35$mm thick with two $4.33$kg weights attached at midspan. The steel plate is placed on pin supports at $30$cm to the both ends, and a hydraulic jack located below the bridge applies an adjustable, horizontal, post-tensioning force. The post-tensioning system allows for control over the stiffness of the model bridge. As seen in Figure \ref{fig:actualsetup}, mobile sensors traversed the bridge using a pulley system. An AppliedMotion STAC6-Si motor towed mobile sensors on four different lanes. The application Si Programmer was used for motion planning and scheduling.\par

Mobile sensors comprised of carts made from K'NEX with rubber bands binding a smartphone to each cart. The phones used the application Sensor Kinetics Pro \citep{SensorKineticsPro} to collect data at the maximum sampling rate ($100$ Hz) with no signal filtering. During testing, six iPhones ranging in model from iPhone6 to iPhone10 were used. Although the application set the sampling rate to 100 Hz, the actual sampling rate varied between phones. Each test required a lab assistant applying moderate-sized, random impulses along the bridge as the cart traveled down the bridge. The goal of the excitation was to simulate an ambient random loads. The cart motion was planned to scan the bridge one way, then pause, and return to the starting position. Therefore, each test consisted of two scans of the bridge by each mobile phone. After collecting a large pool of one-way scans, the data passed through the pipeline shown in Figure \ref{fig:pipeline}. \par

In addition to the mobile sensing tests, the beam was monitored using stationary sensors to create a baseline for validation. Two separate stationary sensing tests were completed. For the first test, five smartphones evenly spaced in the longitudinal direction collected data and then the phones were rearranged to capture torsional modes along two parallel lanes in the transverse direction. The signals were trimmed to 30-second segments for analysis. Finally, the data channels were processed using SMIT \citep{chang2012modal}, a Matlab-based modal identification software to identify natural frequencies and modes shapes. The software allows for selecting a desired SID method among multiple common algorithms. In this study, the ERA-NEXT-AVG method \citep{chang2012modified} was used for processing the stationary data. The resulting frequencies are found in Table \ref{table:MediumSpeedAccuracy} and the natural mode shapes are presented as the baseline in multiple figures (e.g., Figures \ref{fig:AvgCWTMediumSpeed} and \ref{fig:MediumSpeedModeShapes}) and used for MAC value calculations.\par

\subsection{Mobile scans with constant speed}

\begin{figure*}[!h]
    \centering
    \begin{subfigure}{.5\textwidth}
    \centering
    \includegraphics[width=50mm]{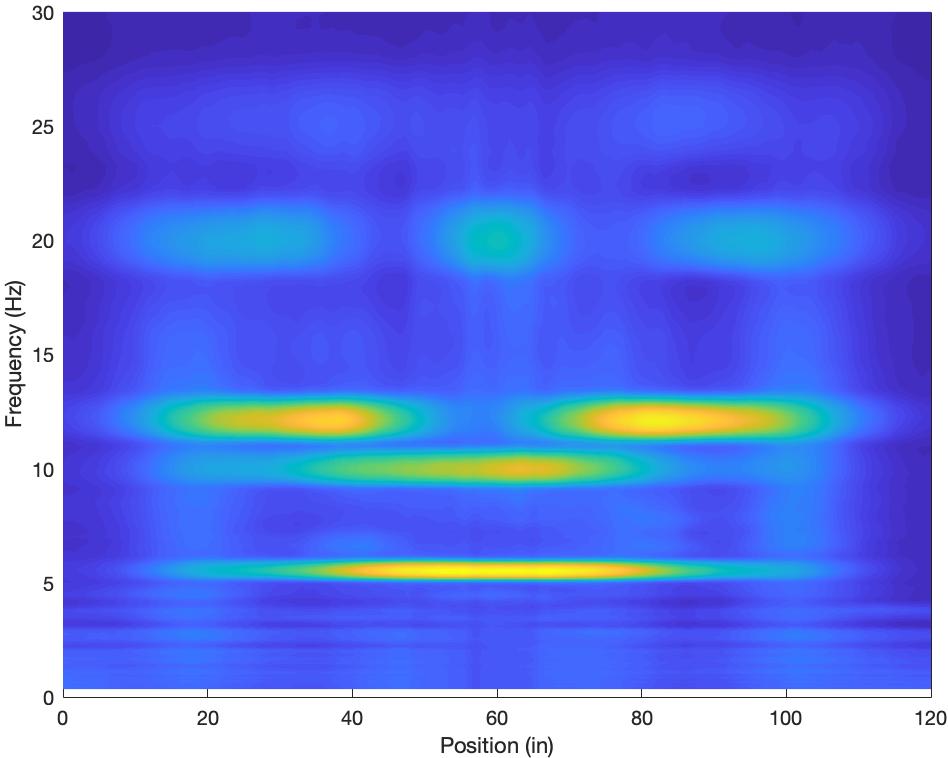}
    \end{subfigure}%
    \begin{subfigure}{.5\textwidth}
    \centering
    \includegraphics[width=60mm]{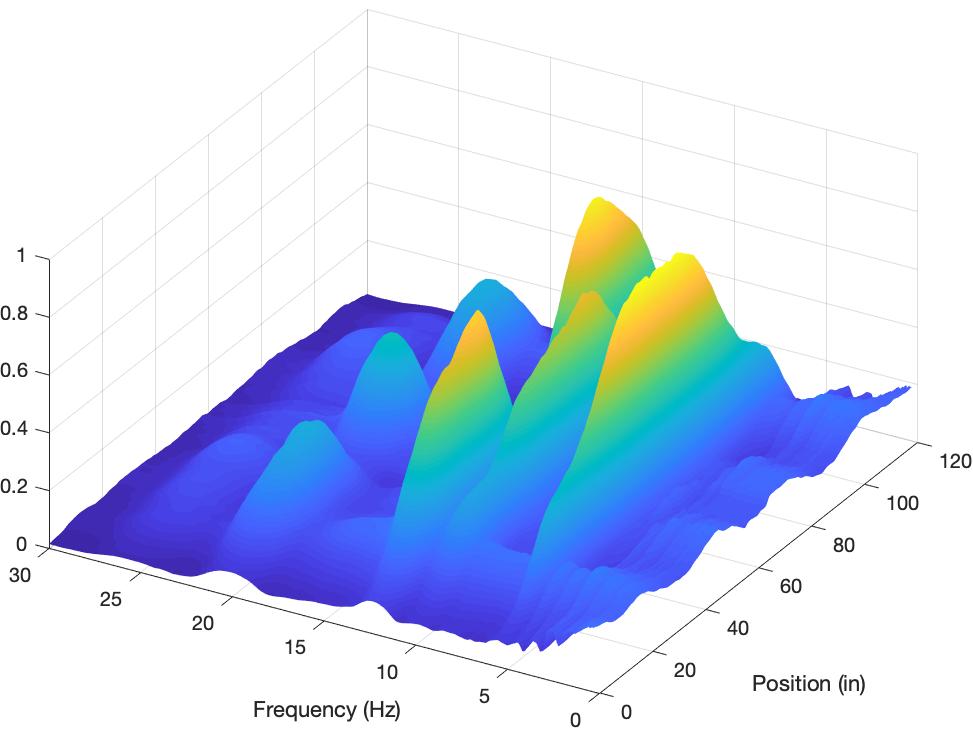}
    \end{subfigure}
    \caption{Aggregated CWTs of 240 one-way scans on the edge lane of the model bridge at medium speed}
    \label{fig:AvgCWTMediumSpeed}
\end{figure*}

\begin{figure*}[!h]
    \centering
    \begin{subfigure}[t]{1.0\textwidth}
    \centering
    \includegraphics[width=160mm]{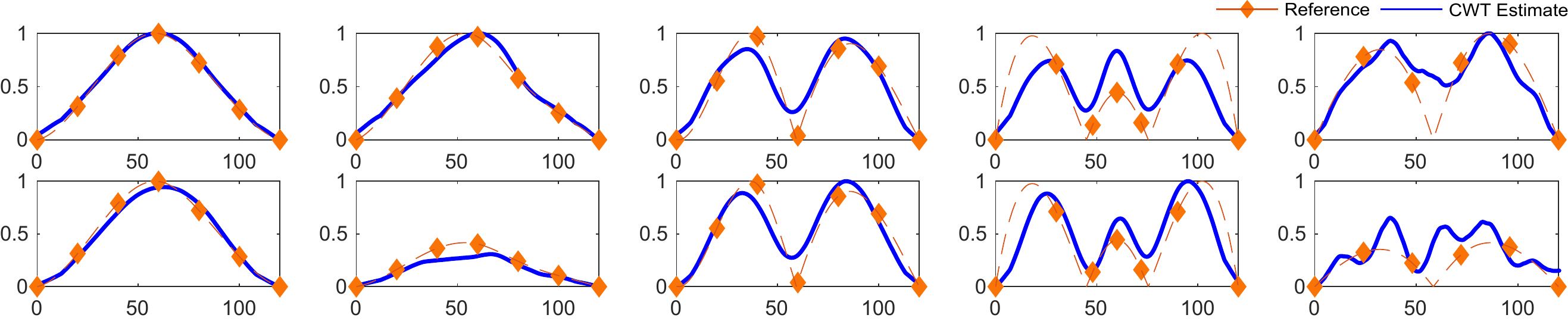}
    \end{subfigure}
    \begin{subfigure}[t]{1.0\textwidth}
    \centering
    \includegraphics[width=160mm]{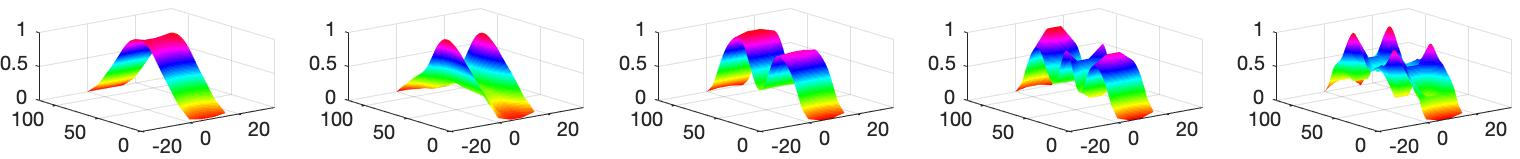}
    \end{subfigure}
    \caption{Three vertical and two torsional absolute mode shapes from 240 scans at medium speed on the edge and middle lane of model bridge.}
    \label{fig:MediumSpeedModeShapes}
\end{figure*}

In the first case, the test includes aggregation of mobile sensors with constant speed to simplify and validate that the proposed pipeline functions properly. In order to achieve this, 240 one-way scans of the bridge with the smartphone-equipped carts were performed. The speed of the sensors was set to $11.38$cm per second (one way scan was $26.4$sec). Aggregating the scans using the pipeline, the test yields the results shown in Figure \ref{fig:AvgCWTMediumSpeed}. This figure displays the aggregated CWT maps in 2D and 3D. Within these maps, the ridges are then picked automatically by a pick-peaking algorithm. The section-cuts of the maps at these peaks are presented in Figure \ref{fig:MediumSpeedModeShapes}. Through aggregation, the undesired random noise in single scans diminishes, leaving the consistently occurring mode shapes. Note that CWT calculates the absolute amplitudes of instantaneous frequencies. In this case, phase information is unavailable since the data sets are not recorded simultaneously. Therefore, the resulting ridges of the CWT map represent the absolute mode shapes of the bridge. \par

\begin{table*}[!h]
\centering
\caption{Identification accuracy measures for the single speed case. Est. stands for estimation.}

\begin{tabular}{@{}lccccc@{}}
\toprule
       & \textit{Mode 1} & \textit{Mode 2} & \textit{Mode 3} & \textit{Mode 4} & \textit{Mode 5} \\ \midrule
Reference freq. [Hz]        & 5.51        & 9.93        & 12.34       & 20.26       & 24.99       \\
CMICW freq. [Hz]              & 5.57        & 10.07       & 12.32       & 20.13       & 25.36    \\
Est. error [\%]              & 1.09        & 1.41       & 0.16       & 0.64       & 1.48   \\
Edge lane, MAC [\%]   & 99.81      & 99.19      & 95.60      & 90.16      & 97.23      \\
Mid lane, MAC [\%] & 99.43      & 99.26      & 94.69      & 98.67      & 82.64           \\ \bottomrule
\end{tabular}
\label{table:MediumSpeedAccuracy}
\end{table*}

The accuracy of identified modal parameters is presented in Table \ref{table:MediumSpeedAccuracy}. The MAC values for the first three modes are above $0.95$ (first two modes above $0.99$) when compared to the fixed sensor baseline, indicating high accuracy in the estimation at lower modes. The accuracy of the estimated mode shapes at two higher modes is also fairly high. Additionally, all the identified frequencies are within $1.5\%$ of the fixed sensor reference found in Table \ref{table:MediumSpeedAccuracy}. These results confirm that CMICW functions as expected and yields high accuracy.\par

\subsection{Speed variations effect}

\begin{figure*}[!h]
    \centering
    \includegraphics[width=160mm]{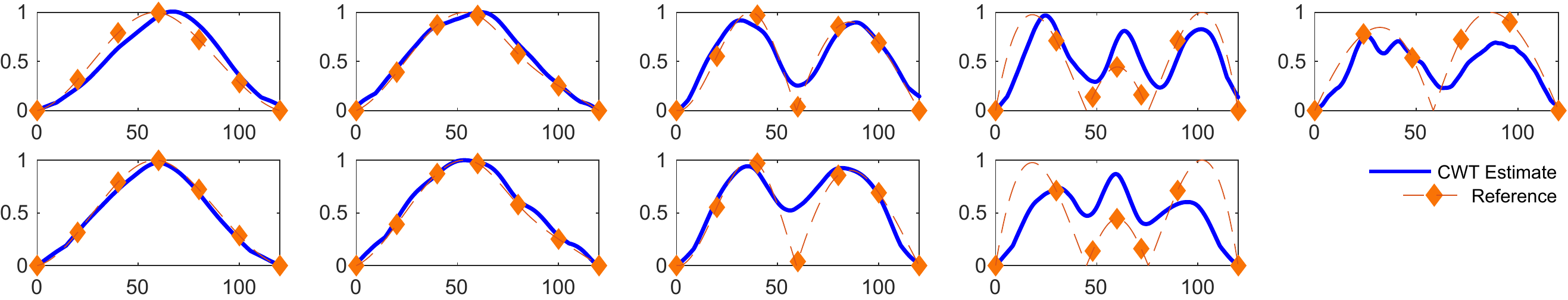}
    \caption{Identified absolute mode shapes with 80 one-way scans using slow sensors (top row) and fast sensors (bottom row). Note that the fifth mode is not identified using fast moving sensors.}
    \label{fig:Multiple Speeds Before Aggregation}
\end{figure*}

\begin{table*}[!h]
\centering
\caption{Identified accuracy measures for different speed cases. Est. stands for estimation.}
\begin{tabular}{@{}lccccc@{}}
\toprule
       & \textit{Mode 1} & \textit{Mode 2} & \textit{Mode 3} & \textit{Mode 4} & \textit{Mode 5} \\ \midrule
Reference freq. [Hz]        & 5.51        & 9.93        & 12.34       & 20.26       & 24.99       \\
Slow CMICW freq. [Hz]   & 5.49      & 10.21      & 12.32      & 20.13      & 25.00      \\
Fast CMICW freq. [Hz] & 5.57      & 10.07      & 11.79      & 19.84      & -      \\
Est. error, Slow [\%]   & 0.36        & 2.82       & 0.16       & 0.64       & 0.04   \\
Est. error, Fast [\%]   & 1.09        & 1.41       & 4.46       & 2.07       & -   \\
Edge lane MAC, Slow [\%]   & 97.55      & 99.59      & 95.87      & 85.91      & 94.53      \\
Edge lane MAC, Medium [\%] & 99.41      & 98.97      & 95.22      & 91.45      & 96.79      \\
Edge lane MAC, Fast [\%]   & 99.87      & 99.78      & 90.12      & 82.47      & -           \\ \bottomrule
\end{tabular}
\label{table:SpeedTestAccuracy}
\end{table*}

The previous experiment is limited to cases in which all scanning carriers retained constant speed. To generalize the application, two additional pools of scans containing (a) 80 slower scans ($9.50$cm per second) and (b) 80 faster scans ($14.53$cm per second) were collected. Three datasets (slow, medium, and fast speeds) are then processed using CMICW (shown in Figure \ref{fig:pipeline}) and results are compared in Figure \ref{fig:Multiple Speeds Before Aggregation} as well as Table \ref{table:SpeedTestAccuracy}. Plot in Figure \ref{fig:Multiple Speeds Before Aggregation} present the identified absolute mode shapes from slow and fast datasets. All five modes were identified accurately in the case of slow speeds. However, with carts moving faster the accuracy of higher modes rapidly decreases. Examining the noise beds - detectable from the local minima of the identified modes at the valleys of the mode shapes - are noticeably higher in the fast speed case (e.g., in the third mode). In other words, as the speed increases, the accuracy of sharp curvature changes in the mode shapes reduce significantly. This can be explained by the fact that faster carriers noticeably magnify noises (e.g., road bumps) compared to slower counterparts. This observation will be further discussed in the following sections and is consistent with existing literature \citep{yang2009extracting}. 

\begin{figure*}[!ht]
    \centering
    \includegraphics[width=160mm]{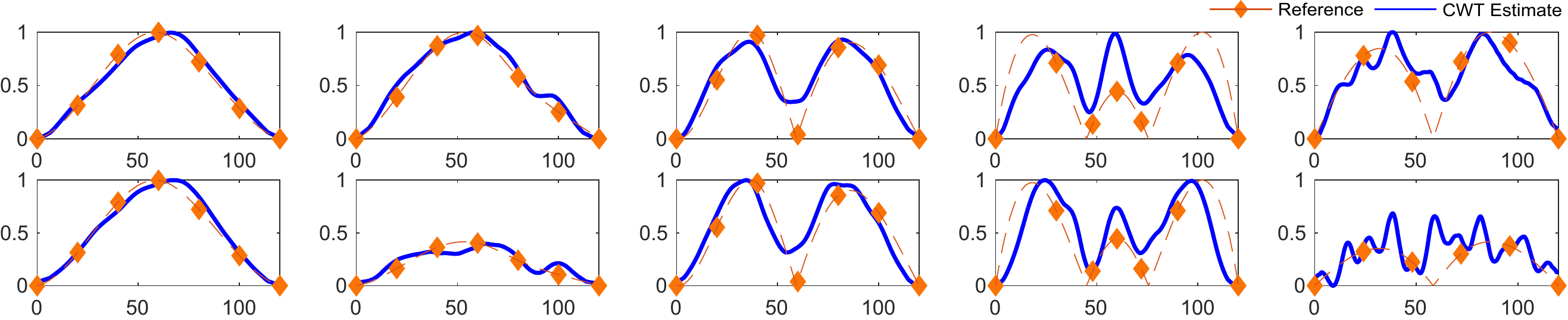}
    \caption{Identified absolute mode shapes through aggregating 80 slow, medium, and fast signals (240 one-way scans total). Top and bottom rows show the edge and middle lane results, respectively.}
    \label{fig:AggregatedResults}
\end{figure*}

\begin{table}[!h]
\centering
\caption{Identification accuracy measures for the aggregated speed case. Est. stands for estimation.}
\begin{tabular}{@{}lccccc@{}}
\toprule
                 & \textit{Mode 1}                & \textit{Mode 2}                & \textit{Mode 3}                & \textit{Mode 4}                & \textit{Mode 5}                \\ \midrule
Reference freq. [Hz]        & 5.51      & 9.93          & 12.34             & 20.26     & 24.99         \\
CMICW freq. [Hz]              & 5.50      & 10            & 12.25    & 20.25     & 25.25         \\
Est. error [\%]              & 0.18        & 0.70       & 0.73       & 0.05       & 1.04   \\
Edge lane, MAC [\%]              & 99.39    & 98.78        & 95.19            & 88.40    & 96.12        \\
Mid lane, MAC [\%]            & 99.17    & 94.93        & 95.01            & 97.94    & 90.73       \\ \bottomrule
\end{tabular}
\label{table:AggTestAccuracy}
\end{table}

\subsection{Mobile scans with varying speeds}\label{Multiple Speed Aggregation}

The objective of this test is to show that the proposed method can incorporate generic car speeds since each car is analyzed independently. A database of 240 CWT scans is composed of 80 one-way scans with slow, medium, and fast speed. The speed-dependent map is interpolated to a preset global space-frequency grid so that it can be aggregated with other scans. Following the same process described in section \ref{sec:method}, the mode shapes are extracted and compared to fixed sensor results in Figure \ref{fig:AggregatedResults}. All five mode shapes that were previously identified in the slow and medium tests are successfully identified in this random aggregation case as well. The aggregated results from different speeds show a comparable accuracy with respect to the constant medium speed test, which has the same sample size. This trial demonstrates that CMICW is applicable for mobile sensors with generalized speeds. Note that the sample pool using which these results are derived is still quite small compared to the data collected on real-world bridges. Bridges that serve a large volume of vehicles daily represent a wealth of information that can be captured by mobile smartphones. It is expected that as the sample size grows, the variance of the estimated modal properties decreases. This is discussed in Section IV. 

\begin{table}[!h]
\caption{Mechanical properties of vehicles used in simulation.}
\centering
\begin{tabular}{@{}lccccc@{}}
\toprule
                     & \textit{V1} & \textit{V2} & \textit{V3} & \textit{V4} & \textit{Units} \\ \midrule
Suspension Stiffness & 62.30       & 128.7       & 2.7e5       & 5700        & N/m            \\
Suspension Damping   & 6.0         & 3.86        & 6000        & 290         & Ns/m           \\
Sprung Mass          & 1           & 1           & 3400        & 466.5       & Kg             \\
Unsprung Mass        & 0.15        & 0.162       & 350         & 49.8        & Kg             \\
Tire Stiffness       & 653         & 643         & 9.5e5       & 1.35e5      & N/m            \\
Tire Damping         & 0           & 0           & 300         & 1400        & Ns/m           \\ \bottomrule
\end{tabular}

\label{table:CarMechanicalProperties}
\end{table}

\subsection{Incorporating suspension effect - a hybrid simulation}\label{Hybrid Simulation}

\begin{figure*}
    \centering
    \includegraphics[width=100mm]{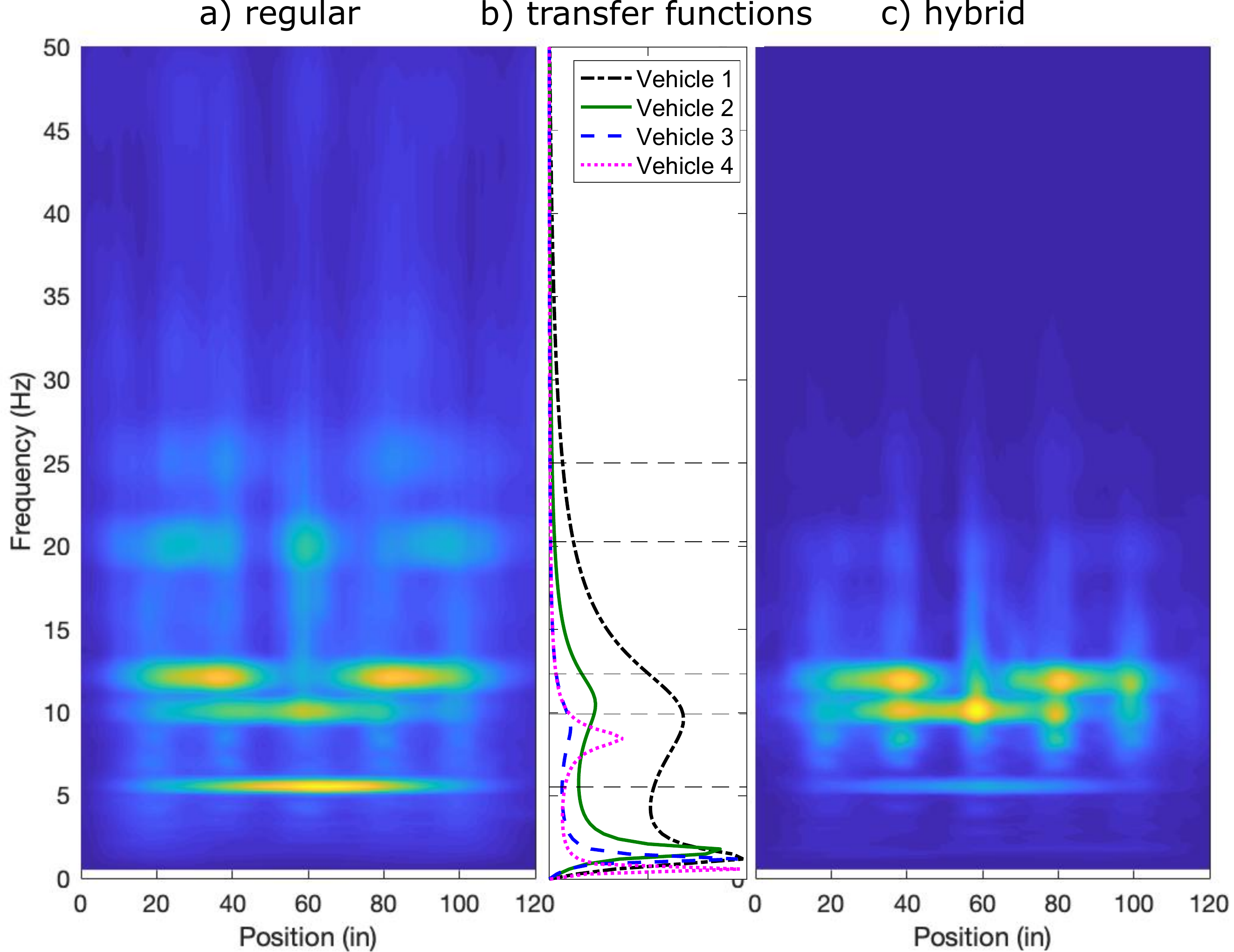}
    \caption{CWT map before and after the hybrid simulation. From (b), higher frequency contents of the bridge are filtered out by the low amplitude tails of the transfer functions, resulting no absolute bridge mode extraction within this high frequency range in (c). This implies that sensing vehicles have to be chosen carefully in order to observe desired frequency range.}
    \label{fig:TransferFunctions}
\end{figure*}

The mobile carts used in the experiments were  stiff; in a real-world scenario with a moving vehicle, the measured accelerations would be affected by the suspension system. This section describes a technique to simulate smartphone-in-vehicle scenarios using a quarter car suspension model to consider a more realistic dynamical response  \citep{malekjafarian2014identification}. The process is as described previously with a new step preceding modal identification: the experimental mobile sensor measurements are fed into a vehicle suspension model. Each individual signal is processed through a linear state-space model which corresponds to the mechanical properties of a quarter-car. To consider the case of multiple vehicle sources, the vehicle properties were selected randomly from the candidates presented in Table \ref{table:CarMechanicalProperties} \citep{sun2001modeling,bogsjo2012models,florin2013passive,gillespie1985measuring}. With this model, both vehicle speeds and suspension systems can be randomized to better simulate real traffic on a bridge. The transfer functions of the four cars are found in Figure \ref{fig:TransferFunctions}. Note that the candidate car types were adopted from other studies, mimicking real-world vehicles in their dynamical properties. Due to the ride comfort concerns, the fundamental mode of vehicle suspensions is around $1.0$ Hz, and the highly damped second natural frequency is approximately $10.0$ Hz. \par

\begin{figure*}
    \centering
    \includegraphics[width=160mm]{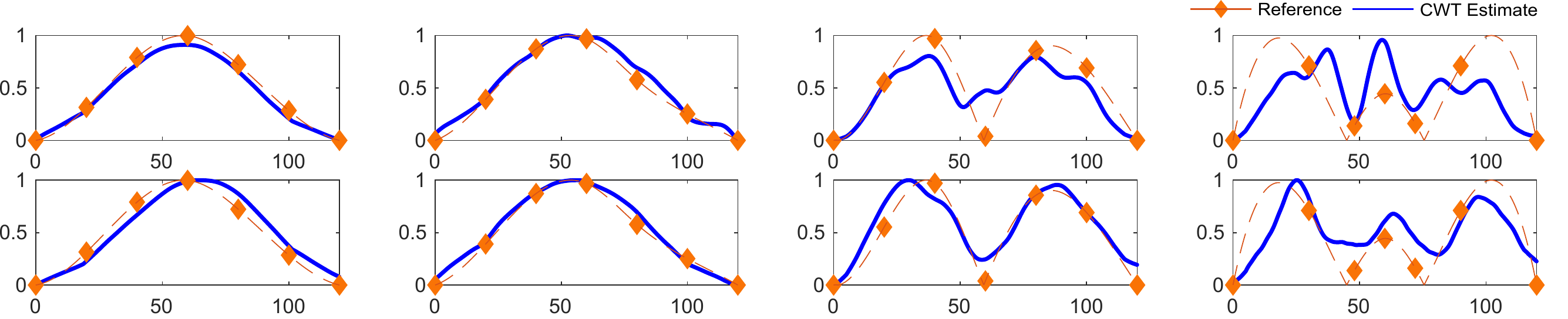}
    \caption{Identified absolute mode shapes from hybrid simulation using only slow moving sensors (top row) and aggregation of different speeds (bottom row).}
    \label{fig:HySimModeShapes}
\end{figure*}

Figure \ref{fig:HySimModeShapes} shows the absolute mode shapes identified by aggregating 80 constant speed slow scans compared to aggregating 240 one-way scans with mix speed aggregation, as seen in Section \ref{Multiple Speed Aggregation}. Additionally, the post-process for both tests remained the same for a fair comparison. Results are consistent with previous experiments. The first three modes are found with high accuracy as presented in Table \ref{table:HybSimAccuracy}. Due to the shape of the transfer functions, certain ranges of the bridge frequency contents are amplified and others diminish substantially, as depicted in Figure \ref{fig:TransferFunctions} (e.g., the fifth mode, at 25 Hz, has nearly vanished). Overall, this effect reduced the contribution of higher modes. In addition, Figure \ref{fig:TransferFunctions}, includes a vertical pattern at five different locations, which is not present in Figure \ref{fig:AvgCWTMediumSpeed}(a). This pattern corresponds to noise generated by the bumps in the experimental setup. Because different speeds have been aggregated, the bump-induced impulses affect the entire frequency spectrum. When the vehicle speed is slow, the effect of these impulses on mode shape estimation is less significant. The result is an accuracy that is comparable with an analysis using only one-third the sample size. To mitigate the ``bump effects'', an average of the noise bed over the CWT map is removed from the aggregated map. \par

\begin{table}[]
\centering
\caption{Identification accuracy measures for the hybrid simulation. Agg. and est. stand for aggregated and estimation, respectively.}
\begin{tabular}{@{}lcccc@{}}
\toprule
                  & \textit{Mode 1} & \textit{Mode 2} & \textit{Mode 3} & \textit{Mode 4} \\ \midrule
Reference freq. [Hz]        & 5.51        & 9.93        & 12.34       & 20.26       \\
CMICW slow freq. [Hz]     & 5.72        & 9.67       & 12.15       & 20.31       \\
CMICW agg. freq. [Hz]   & 5.72        & 10.42       & 13.38       & 19.81       \\
Est. error, Slow [\%]   & 3.81        & 2.62       & 1.54       & 0.25    \\
Est. error, Agg. [\%]   & 3.81        & 4.93       & 8.43      & 2.22     \\
Edge lane MAC, Slow [\%]        & 99.52      & 99.35      & 94.16      & 87.76      \\
Edge lane MAC, Agg. [\%]  & 99.86      & 99.31      & 89.94      & 78.35      \\ \bottomrule
\end{tabular}
\label{table:HybSimAccuracy}
\end{table}

\section{Discussion and Conclusion}\label{sec:dicussion}

\subsection{Statistical analysis}

\begin{figure}[h]
    \centering
    \includegraphics[width=85mm]{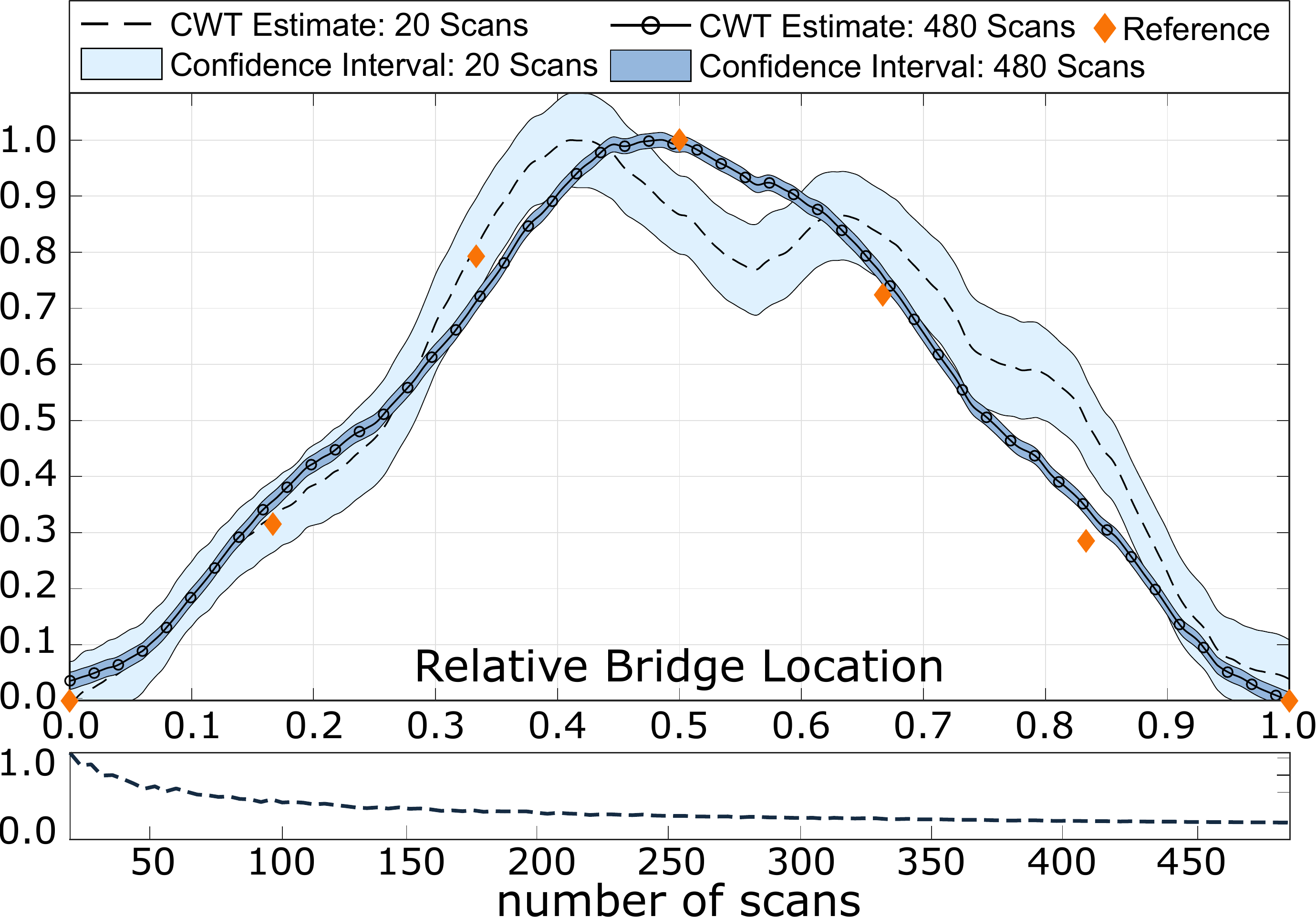}
    \caption{Identified fundamental mode using 20 scans versus 480 scans: the confidence interval of the identified points has significantly narrowed by crowdsourcing more data. Below the confidence interval width vs. number of aggregated scans is plotted.}
    \label{fig:AccuracyVsSampleSize}
\end{figure}

CMICW can operate exclusively from crowdsourcing. In the previous sections, it was shown that when the sample size is large, more accurate results are expected. However, the confidence of estimations with respect to the sample size has not been investigated. Figure \ref{fig:AccuracyVsSampleSize} presents the identification accuracy of the fundamental mode using 20 samples versus 480 samples (selected from medium speed scans). The lines show the average of mode shapes from all individual scans and the shady area around centerlines indicate the $95\%$ confidence intervals in the estimations. The figure demonstrates that by increasing the number of samples, not only the accuracy of the estimations increase (i.e., shape of the mode), but also the confidence of estimations. In the lower part of the figure, the trend of confidence interval width versus the sample size is presented, consistent with this finding. The fact that the mode shape identification result eventually becomes more reliable (e.g., $30\%$ reduction in CI when number of scans grow from 50 to 100) show the promising strength of this crowdsourcing-based method. \par

\begin{figure*}[h]
    \centering
    \begin{subfigure}[t]{0.5\textwidth}
    \centering
    \includegraphics[width=90mm]{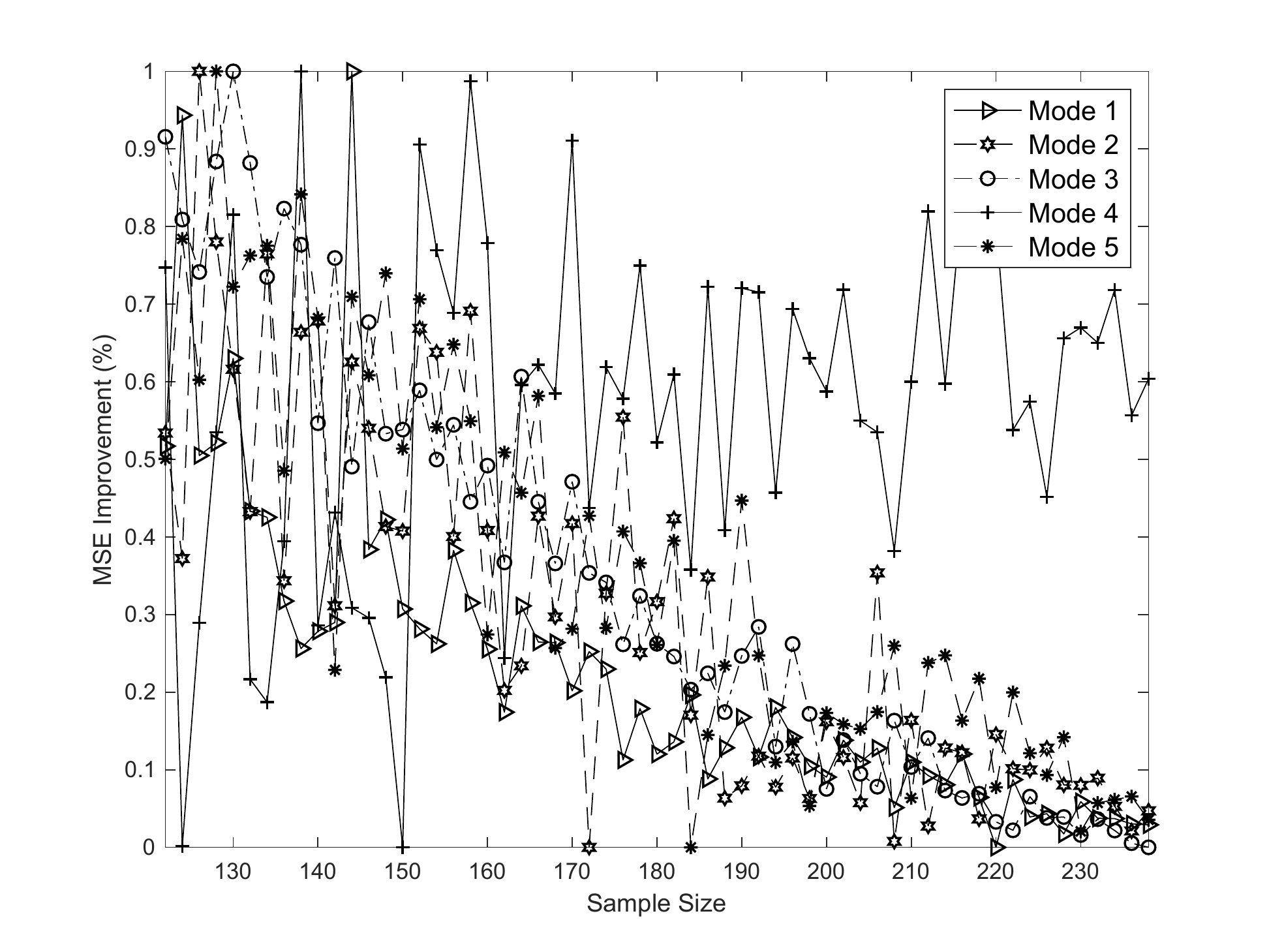}
    \caption{MSE trend on 3D modes}
    \end{subfigure}%
    \begin{subfigure}[t]{0.5\textwidth}
    \centering
    \includegraphics[width=90mm]{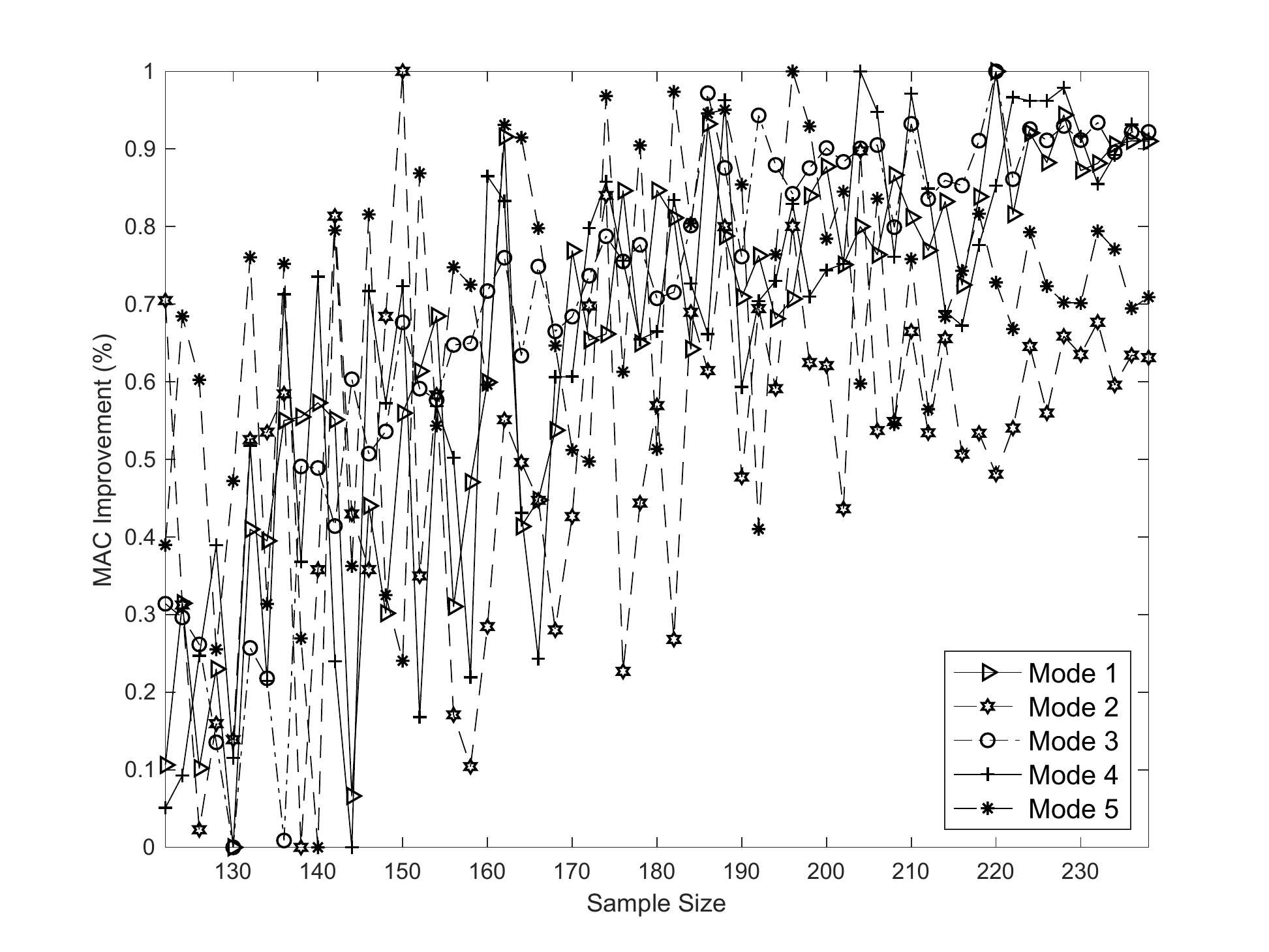}
    \caption{MAC trend on edge lane}
    \end{subfigure}
    \caption{Accuracy measures vs. sample size. Both figures confirm that as the number of scans increases, the error reduces and the accuracy also improves.}
    \label{fig:discussion_improvements}
\end{figure*}

To generalize the scalability of the method, in Figure \ref{fig:discussion_improvements}, the trends of the identified modes' MSE and MAC values are analyzed with respect to the sample size. For maximum fairness of the trial, $100$ random sample sets are picked for each sample size, and results are averaged. In all cases, as the size of the aggregation pool increases, it is more likely to predict modal properties accurately. Note that in some cases, one may pick a very nice subset of the scans, using which the estimation outperforms the estimations with more samples. This supports the idea that by aggregation of larger pool of data, the reliability of estimations increases, because they gradually become less and less sensitive to the quality of individual samples. \par

\begin{figure}[h]
    \centering
    \includegraphics[width=75mm]{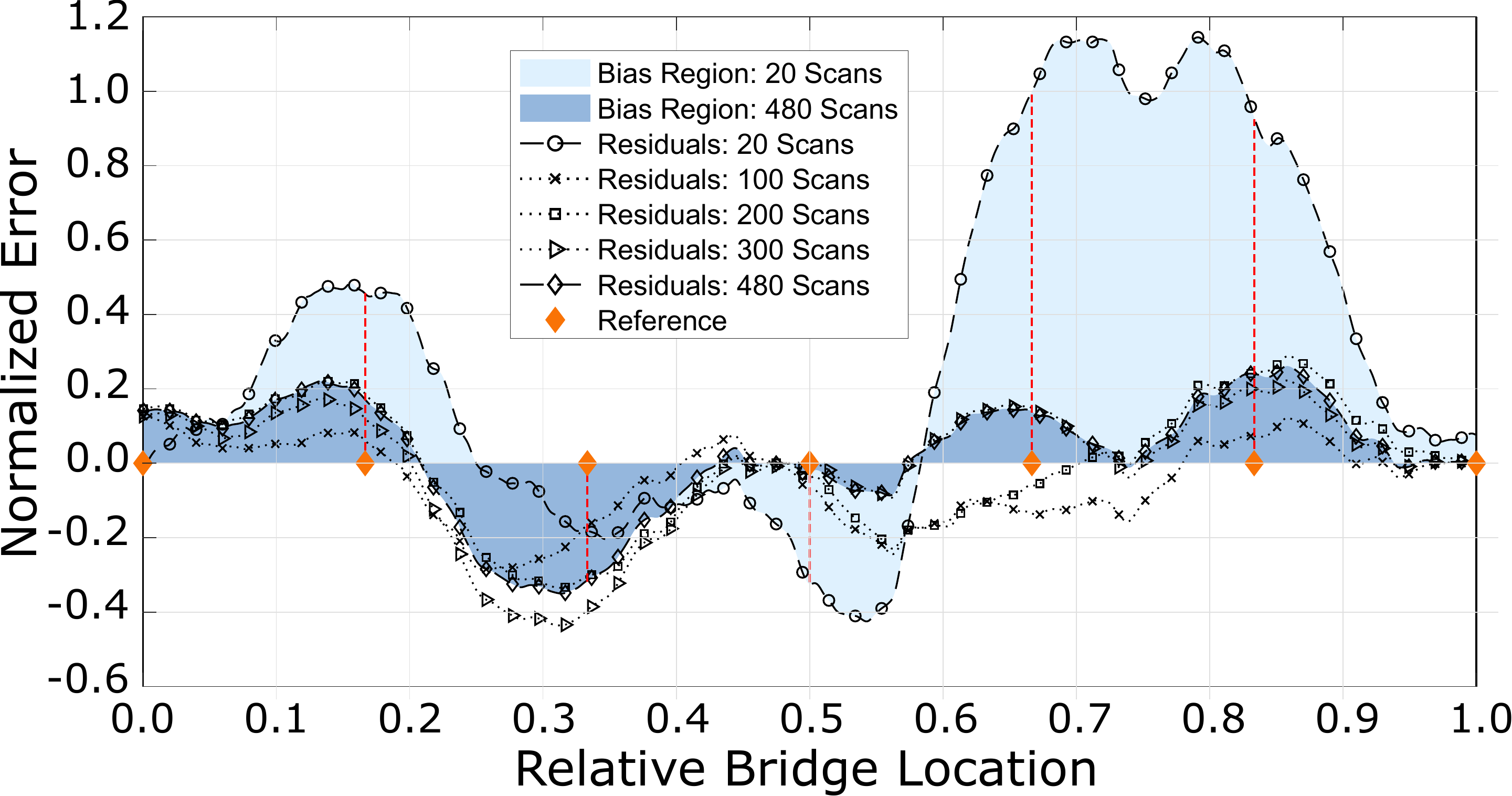}
    \caption{Unreducible bias analysis based on the first mode shape. As the sample size grows, the error at stationary locations reduces. However, the error does not reach zero due to the presence of consistent physical obstacles (here, transversal wires).}
    \label{fig:biasAnalysis}
\end{figure}

CMICW is prone to intrinsic biases when it comes to the spatial information. These biases were visible in Figure \ref{fig:AccuracyVsSampleSize} where the reference points (found using fixed sensors) fall outside the band of confidence interval (e.g., with 480 scans). In other words, the plot shows that in some locations, the aggregation yields a confident estimation that is slightly inaccurate (e.g., the reference point at relative location $0.83$ in Figure \ref{fig:AccuracyVsSampleSize}). In this context, the bias is explained as an unreducible error in the mode shape amplitudes caused by consistent physical obstacles for a smooth sensing, such as expansion joints and speed bumps (note that road irregularity is not a consistent obstacle since it may or may not appear in a single scan). These obstacles (among other variables) introduce an undesired content to every scan, causing a bias in the final amplitude of the modal ridges at those location. Since this effect happens in every scan, it is not reducible by adding more scans to the pool. To better examine these, Figure \ref{fig:biasAnalysis} shows errors in identified mode shapes via different sample sizes. The red lines indicate the errors in the estimated fundamental absolute mode shape when compared with the stationary sensors result. The bias between small and large sample sizes is reduced considerably (in particular, $92\%$ bias reduction between 20 scans and 480 scans), nevertheless, the error is still present in the case with the larger sample size. \par

\begin{figure*}[h]
    \centering
    \includegraphics[width=180mm]{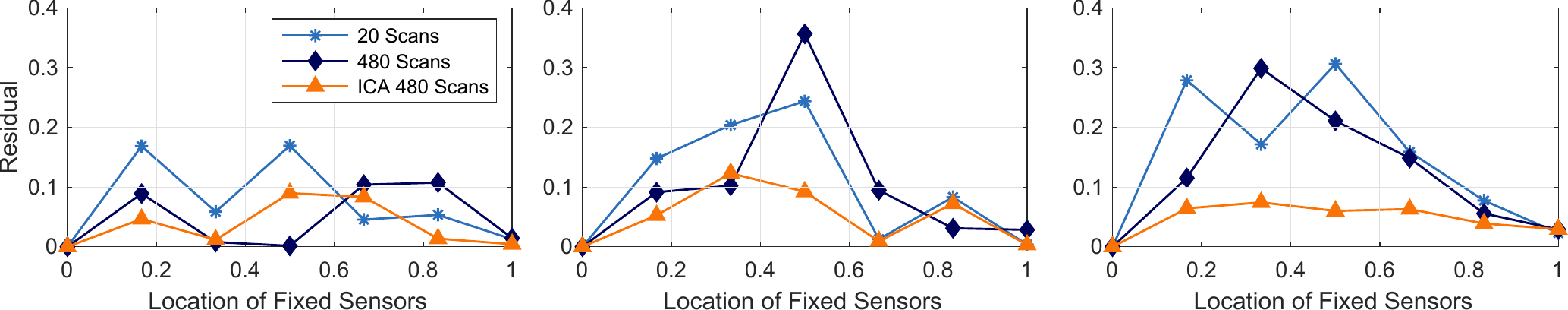}
    \caption{Identified mode shape error before and after applying ICA for separation of the effects of consistent obstacles (here, wire bumps) for the first three modes. For higher modes, ICA dramatically helped to reduce the error. In particular, the accuracy of the third mode (right) has improved by 70\% after ICA post-processing.}
    \label{fig:ICA_effect}
\end{figure*}

In our experimental setup, there were transversal wires that act as speed bumps and joints, which can explain the unreducible errors in those locations. To remove this undesired content, blind source separation techniques can be applied as proposed in \cite{eshkevari2020deconv} and \cite{eshkevari2019bridge}. In this approach, given multiple scans over a certain portion of the bridge, algorithms such as ICA or SOBI can extract the common content - consistent road bumps or joints - as one of the sources. Once it is extracted, mobile scans can be filtered and the aggregation will be performed on the filtered signals. The method is implemented here and the final effect on the unreducible error is presented in Figure \ref{fig:ICA_effect}. In this figure, a comparison between error values before and after applying ICA demonstrate that the decontamination technique enhanced the performance of the algorithm. In particular, the effect on higher modes is more significant. 

\subsection{Final remarks}

This paper presented and experimentally validated a novel crowdsensing approach for bridge health monitoring using smartphones. In particular, CMICW method is able to estimate natural frequencies and absolute mode shapes. The method aggregates spatio-temporal maps of the individual signals collected by mobile smartphones to estimate the modal properties of a bridge. In this study, we showed that the approach has the following advantages:

\begin{itemize}[]
    \item CMICW is a fully crowdsourcing based approach using widely available smartphones for bridge modal identification. 
    \item This method does not require synchronized or simultaneous sensors, which maximizes its implementation. 
    \item The MAC accuracy of identified mode shapes enhances as more data become available. In addition, by aggregating more data, the estimations become more reliable (i.e., narrower CI). 
    \item Vehicle suspension results in a loss of higher modes; however, as long as the suspension properties of the vehicles that are used in a large dataset are random, the frequency contents of the suspension itself vanish through the aggregation process and the lower modes are accurately identified. 
    \item By aggregation of results from different lanes of a bridge, a 3D representation of the identified modes is achievable. 
\end{itemize}

CMICW was validated using an experimental setup. In summary, the first five natural modes of the beam setup were targeted for the modal identification (three vertical and two torsional modes). Three test cases were conducted: (a) 240 medium speed scans, (b) 80 scans with fast and slow speeds (each), and (c) hybrid simulation of vehicle suspension. By aggregation of the samples with constant speeds, we were able to identify all five modes for the slow and medium speed cases and first four modes with the fast speed case (e.g., first two natural modes with MAC $99\%$ and frequency estimations within $1.4\%$ error). In the second aggregation strategy, 80 scans from each speed were picked randomly to simulate a more realistic sample set. The aggregation resulted all five modes with MAC estimation of $>95\%$ for the first three modes and frequency estimations within $1.0\%$ error. In the final aggregation strategy, the same sample set described previously was passed through a random group of quarter-car suspension models in order to simulate vehicle manipulation of signals. From this hybrid simulation, the first four modes were identified (e.g., the first three modes yielded MAC values $>90\%$). The fifth mode was not identifiable due to its overlap with the low amplitude range of the vehicle transfer function. The hybrid simulation also was repeated using only $80$ slow scans, that resulted in higher estimation accuracy (e.g., the first three modes yielded MAC values $>94\%$). \par
CMICW is a promising solution for large-scale, fully data-driven, crowdsourcing-based method for the structural health monitoring of bridges. 

\section*{Acknowledgment}

Research funding is partially provided by the National Science Foundation through Grant No. CMMI-1351537 by the Hazard Mitigation and Structural Engineering program, and by a grant from the Commonwealth of Pennsylvania, Department of Community and Economic Development, through the Pennsylvania Infrastructure Technology Alliance (PITA). The authors would like to thank Anas S.p.A, Allianz, Brose, Cisco, Dover Corporation, Ford, the Amsterdam Institute for Advanced Metropolitan Solutions, the Fraunhofer Institute, the Kuwait-MIT Center for Natural Resources and the Environment, LabCampus, RATP, Singapore-MIT Alliance for Research and Technology (SMART), SNCF Gares \& Connexions, UBER, and all the members of the MIT Senseable City Lab Consortium for supporting this research.

\bibliographystyle{unsrtnat}
\bibliography{Bibliography}

\end{document}